# Active microrheology of *Chaetopterus* mucus determines three intrinsic lengthscales that govern material properties

W.J. Weigand[+], A. Messmore[+], D.D. Deheyn[+++], A. Morales-Sanz[++], D.L. Blair[++], J.S. Urbach[++], R.M. Robertson-Anderson[+*]

[+]*Department of Physics, University of San Diego, San Diego, CA 92110*
[++]*Department of Physics and Institute for Soft Matter Synthesis and Metrology, Georgetown University, Washington, DC, 20057*
[+++] *Marine Biology Research Division, Scripps Institution of Oceanography, La Jolla, CA 92037*

**Abstract:**

We characterize the scale-dependent rheological properties of mucus from the *Chaetopterus* sea annelid and determine the intrinsic lengthscales controlling distinct rheological and structural regimes. The mucus produced by this ubiquitous filter feeder serves a host of roles including filtration, protection and trapping nutrients. The ease of clean mucus extraction coupled with similarities to human mucus rheology also make *Chaetopterus* mucus a potential model system for elucidating human mucus mechanics. We use optically trapped microsphere probes, 2-10 µm in size, to induce oscillatory microscopic strains and measure the stress response of the mucus as a function of oscillation amplitude and frequency. We show that the viscoelastic properties are highly dependent on the scale ($l$) of the strain with three distinct regimes emerging: microscale: $l_1$ ≤ 4 µm, mesoscale: $l_2$ ≈ 4-10 µm, and macroscale: $l_3$ ≥ 10 µm. While the mucus response is similar to water for $l_1$ indicating that oscillating probes rarely contact the mucus mesh, for $l_2$ the response is distinctly more viscous and independent of probe size, demonstrating that the mucus network behaves as a continuum. However, this principally viscous mesoscale response at $l_2$ is distinct from the largely elastic macroscopic mucus response. Only for $l_3$ does the response mimic the macroscopic elasticity, with rigid constraints strongly resisting microsphere displacement. Our results demonstrate that a uniform biopolymer mesh model for mucus with a single lengthscale modulating the crossover from water-like to elastic response is too simplistic. Rather, the mucus responds as a hierarchical network with a loose microscopic mesh controlling mechanics for $l_2$, coupled with a mesoscale rigid scaffold responsible for the macroscopic gel-like mechanics beyond $l_3$. Our results shed important new light onto the design of drug delivery platforms, preventing pathogen penetration, and improving filtration, coating and clearance capabilities of mucus.

**Introduction:**

Mucus is a ubiquitous viscoelastic gel present in all animals that protects against foreign pathogens (1-3), coats organs, and aids in digestion (4, 5), while circulating oxygen and allowing the passage of nutrients past the mucus barrier (6). Filter-feeder marine animals, such as the *Chaetopterus* marine worm, also use mucus to trap and filter food, deter prey and build housing tubes (7-9). The myriad of essential mechanical roles that mucus plays are critically dependent on its rheological properties. In fact, the most debilitating diseases that target mucus, such as cystic fibrosis, alter its rheology. Further, the multifunctional role of mucus, allowing for simultaneous mobility of nutrients at the microscale and coating and lubrication at the macroscale, suggests that the rheological properties vary over different lengthscales, allowing for both fluid-like and elastic responses at different scales.

The multifunctional properties of mucus likely stem from the complex structure of the biopolymer mesh comprising the mucus. Mucus glycoproteins (mucins), DNA, lipids, ions, proteins, cells, cellular debris, and water are all components of mucus, with mucins playing the most significant role in the physical properties of the mucus (10-12). Due to the important and complex role that both mechanics and structure play in mucus function, a number of researchers have investigated the rheological and structural properties of mucus from a range of animals and organs, with an underlying goal of elucidating human mucus properties (13-16). While the molecular structure of mucus in all animals and organs is somewhat similar, different studies have reported wide-ranging results (5, 13, 16-22).

Both macro- and micro- rheological methods have been used to investigate the material properties of mucus (23). Both techniques are able to extract viscoelastic properties including the elastic or storage modulus, $G'$, viscous or loss modulus, $G''$, and viscosity, $\eta$. Macrorheology applies macroscopic strains to a material and measures the bulk stress response, while microrheology uses micron sized probes embedded in the material to determine material properties. Passive microrheology uses the Brownian motion of embedded probes to infer material properties while active techniques use optically or magnetically trapped probes to locally perturb the material and measure the induced molecular-level stress. While microrheology measurements use Stokes-Einstein relations to relate probe motion to material properties, this method relies on the assumption that the material can be treated as a continuum. In this continuum limit, the measured material properties are independent of the size of the probe and are comparable to the bulk measured properties. However, biopolymer networks such as mucus have a range of important intrinsic lengthscales, such as polymer persistence lengths and network mesh sizes, that prevent this continuum limit from being unequivocally valid at all lengthscales. In such networks the continuum limit is generally understood to be achieved for probes larger than the characteristic mesh size of the molecular network. Probes smaller than the mesh size travel largely within the pockets or pores of the molecular mesh and thus principally characterize the Newtonian fluid surrounding the mucus network. However, this naïve assumption has been called into question by a number of studies that have reported larger correlation lengthscales in biopolymer networks beyond the mesh size (24). We previously determined that for entangled networks of DNA, the probe size necessary to access the continuum limit was ~5x the diameter of the entanglement tube in which each DNA molecule is confined (25). Thus, measurements carried out with a range of probe sizes can not only measure rheological properties over a range of lengthscales, but can also quantify the intrinsic correlation lengths or entanglement spacings present in the material.

Macrorheology experiments have reported viscosities of human mucus to be $10^4$-$10^6$x that of water and exhibit shear thinning at high shear rates (26-30). Macroscopic measurements of $G'$ and $G''$ also show that most mucus systems behave as elastic-like gels rather than viscous liquids (17, 31-37). These results are in contrast to several microrheology studies which show that mucus responds as a purely viscous Newtonian fluid (5, 24, 38-40). Passive microrheology studies have focused on identifying the mucus mesh size by determining the probe size below which probe transport mimics that of probes traveling through water, reporting wide ranging results (5, 39, 41-45). Passive microrheology studies on human cervicovaginal mucus reported an average pore size of ~340 nm with a wide range of 50-1800 nm (41), while other groups report a range of 200 – 370 nm (42). Authors suggest that this wide range indicates that the mucins can bundle and create cables that are ~3x thicker than individual mucin fibers (41, 43). In contrast, two studies on human cervical mucus, using FRAP and fluorescence imaging, found that for ~10 nm particles the viscous drag that the probes experienced was similar to that found in water while 180 nm probes are slowed ~100x (39, 44). FRAP has also been used to measure the diffusion of both linear and supercoiled DNAs in bovine cervical mucus and infer a mucus pore size of ~12.5 μm (45), in agreement with pore size estimates of 1 – 20 μm, measured using electron (46) and confocal microscopy (47). Alternatively, electron microscopy measurements on human mucus measure pore sizes of 100 – 380 nm (44).

Active microrheology, which can probe larger lengthscales and exert higher forces than passive techniques, has been used far less to probe mucus systems. Recent active measurements of zebra fish intestinal mucus reported purely viscous response with viscosity ~5x that of water for length scales up to ~10 μm (5). Conversely, active studies on pulmonary mucus reported evidence of a rigid scaffold with a mesh size on the order of microns which authors infer from 1 μm optical trap oscillations in which ~80% of probes could not stay trapped during oscillation (43). Corresponding electron microscopy measurements revealed large heterogeneities in pore size ranging from 500 nm to 10 μm (43).

Here we use active optical tweezers microrheology to characterize both the structure and scale-dependent rheological properties of the mucus produced by the *Chaetopterus* marine worm. Previous macrorheology measurements on this mucus show that the viscous stress response is highly comparable to human mucus (48). Further, the worms excrete ~10-100 μl of mucus upon modest agitation, allowing for reliably clean mucus samples to be collected. Provided the similarities to human mucus and the ease of clean extraction, this mucus represents a potential model system that can provide insights into human mucus structure and rheology.

We sinusoidally drive microspheres of various sizes (2-10 μm) and measure the resulting force acting on the bead. From these measurements we determine the storage modulus, $G'$, loss modulus, $G''$, and the complex viscosity, $\eta$ of the mucus. We find three rheologically distinct lengthscales in the mucus $l_1 < 4$ μm, $l_2$ ~4-10 μm, and $l_3 > 10$ μm. For $l_1$ the stress response is comparable to water as the probes move largely within the pores of the mucus network. For $l_2$, mucus can be treated as a continuum material that exhibits loose entanglements and predominantly viscous response properties. The macroscopic elastic gel properties measured at the macroscale only arise for $l_3$ where a rigid mesh structure forces the probe out of the optical trap. Our results shed important new light onto how mucus is capable of simultaneously performing a range of functions that require different material properties.

**Materials and Methods:**

The *Chaetopterus* marine worms used in this study were collected as previously described (49). They were still inside their tubes and bundles of tubes were then transferred in buckets filled with cold seawater for the transit between the Scripps Institution of Oceanography and the University of San Diego (<10 km). The bundles of tubes were stored at USD in filtered artificial seawater at ~15°C. The mucus extraction method used is similar to that used previously (49). We removed a worm from its housing tube, placed it on a petri dish with a small amount of seawater, and removed the head and parapodia from the rest of the body (Figure 1). Mucus from the head and parapodia was extracted by applying light pressure to the body parts using a syringe and collecting the excreted mucus, which was stored at 4°C for up to 1 hour.

Carboxylate microspheres (beads; Polysciences) of 2, 4.5, 6, and 10 µm are coated with Alexa-Fluor 488 bovine serum albumin (Invitrogen) to prevent nonspecific binding to mucus and visualize the beads. For all experiments, we mixed 3 µL of 0.5% (v/v) beads with 10 µL of mucus and 0.015% Tween-20 (Fisher Scientific) which prevents beads from binding to the mucus and microscope slide. We then injected the mucus mixture into a microchannel comprised of a microscope slide and coverslip separated by two layers of double-sided tape, and sealed the ends with epoxy.

During measurements, individual microspheres were trapped using a custom built optical trap formed by a 1064 nm Nd:YAG fiber laser (Manlight) focused by a 60x 1.4 NA objective (Olympus). The sample was precisely oscillated sinusoidally relative to the trapped bead by using a piezoelectric nanopositioning stage (Mad City Laboratories). A position-sensing detector (PSD, Pacific Silicon Sensors) was used to measure the laser deflection, which is proportional to the force $F$ on the microspheres. The force-detection was calibrated for each bead size using Stokes drag method for a microsphere in water (50).

Following measurements, we fit both the measured force $F$ and stage position $x$ data to sine curves using the least squares method as described previously (51) (Figure 1). The fits are then used to calculate the storage modulus $G' = [|F_{max}|cos(\Delta\phi)]/[|x_{max}|6\pi R]$, the loss modulus, $G'' = [|F_{max}|sin(\Delta\phi)]/[|x_{max}|6\pi R]$, and the complex viscosity, $\eta = (G'^2 + G''^2)^{1/2}/\omega$ where $|F_{max}|$, $|x_{max}|$, $\Delta\phi$ and $R$, are the amplitude of measured force, amplitude of the stage position, phase difference between the $F$ and $x$, and bead radius respectively (51). Five trials, each averaging 10 full oscillations were completed for each frequency, amplitude, and bead size. For each trial, a different bead in a different region of the sample was used. All data shown (Figs 3-7) are averages over three different worms with error bars representing the corresponding standard error. We carried out measurements for oscillation amplitudes of 1 – 8 µm and frequencies of 1 – 15 Hz.

For macrorheology measurements, mucus samples were prepared by removing the head and parapodia from the worm and crushing them in a plastic syringe. The mucus collected from the syringe was mixed with water to produce a 50% v/v solution. We used a rotational rheometer (Anton Paar) equipped with a cone plate tool measuring 25 mm in diameter to exert oscillatory strain on the mucus solution and measure its elastic and loss moduli. The strain on the mucus started at 0.01% and slowly increased to 5%. The oscillatory frequency was kept constant at 1Hz. Macrorheology data shown (Figs 5, 7) is an average of two individual worms and corresponding standard error.

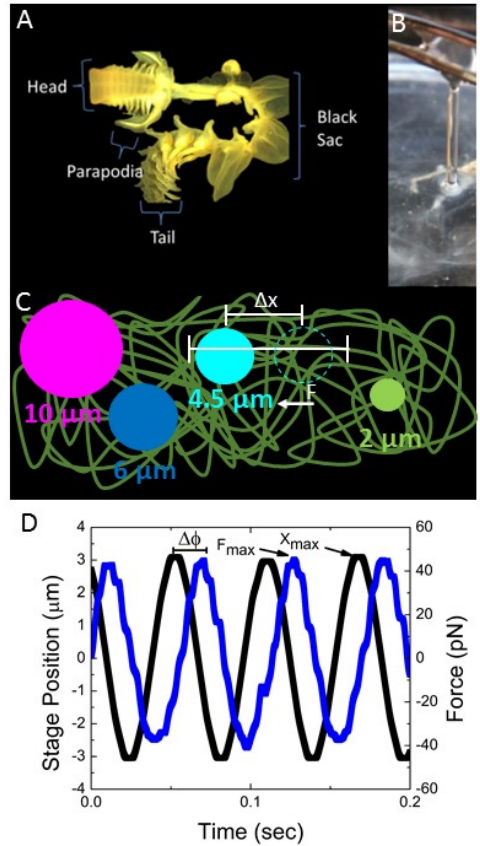

**Figure 1: Schematic of active microrheology measurements of *Chaetopterus* marine worm mucus.** (A) Typical *Chaetopterus* worm used for experiments, with body parts labeled. Mucus is extracted from the head and parapodia. (B) Image of viscoelastic mucus excreted by the worm. (C) Schematic of sample chamber containing microspheres embedded in the mucus. Bead sizes are labeled and colored as they appear in the rest of the figures. Here, the 4.5 μm bead is trapped and sinusoidally oscillated at an amplitude $\Delta x$ while the force $F$ on the bead is measured. (D) Sample force $F$ (blue) and stage position $x$ (black) curves are shown for 6 μm beads. $F_{max}$ is the amplitude of the force exerted by the mucus on the bead, $x_{max}$ is the oscillation amplitude, and $\Delta\phi$ is the phase difference between the force and stage curves.

**Results and Discussion:**

As shown in Figure 2, oscillatory force measurements show a complex dependence on probe size. As described in the Introduction, if the mucus can be treated as a uniform continuum, the measured force $F$ should be proportional to the probe size $R$, resulting in probe-independent material properties (i.e. $G'$, $G''$, $\eta \sim F/R$). For the 4.5 and 6 $\mu$m beads, the force response normalized by the probe size is indeed independent of the probe size indicating that at these lengthscales the biopolymer mesh comprising mucus can be treated as a continuum material. In contrast, the 2 $\mu$m beads measure a ~2x lower normalized force indicating that they interact less with the mesh network. These discrete effects indicate that the transition to the continuum regime is ~4 $\mu$m. This lengthscale is substantially (~5x) larger than previous estimates measured via passive microrheology and microscopy in human and bovine mucus systems (39, 41-44). This finding suggests that the mesh is very loosely entangled so it only contributes appreciably to the force response due to active perturbations for lengthscales ~5x the mesh size. These results are in agreement with our previous results for entangled DNA in which we determined that the transition to continuum occurred at ~5x the entanglement spacing of the network (25). However, mucus mesh size estimates as high as 20 $\mu$m have been made based on microscopy methods and theoretical predictions (45, 48, 49, 52, 53), suggesting that there are perhaps multiple lengthscales that control mechanics of mucus systems.

Indeed, contrary to the typical single fluid model, where the response is independent of bead size above one critical lengthscale, defining the continuum regime (54), we find that for 10 $\mu$m beads the normalized force drops by a factor of ~3 when compared to the 4.5 and 6 $\mu$m beads (Figure 2). Upon closer examination we see that the force oscillation curves have relatively flat or sunken peaks rather than the expected sinusoidal response seen with the other probe sizes (Figure 2b). These small dips in the force peaks demonstrate that the bead is briefly forced out of the trap during oscillation and is picked up again as the stage returns to equilibrium. This forcing is likely due to the probe encountering a more rigid network or structure that is strong enough to force it out of the trap (43). Resulting viscoelastic moduli calculated from the truncated force curves (Figures 3,4,6) will have similarly smaller amplitudes, as these moduli are proportional to $|F_{max}|$. Previous studies in multiple mucus systems, have also shown evidence of a similar rigid scaffold at larger lengthscales, likely due to mucin bundling and crosslinking (41, 43, 55).

These results suggest that the mucus is a "multiscale mesh" with three distinct length scales: $l_1 \leq 4$ $\mu$m, $l_2 \approx 4$-10 $\mu$m, and $l_3 \geq 10$ $\mu$m. To test this hypothesis and characterize the mechanical properties at each of these lengthscales, we turn to the amplitude, frequency and probe size dependence of the resulting viscoelastic properties derived from the measured force response.

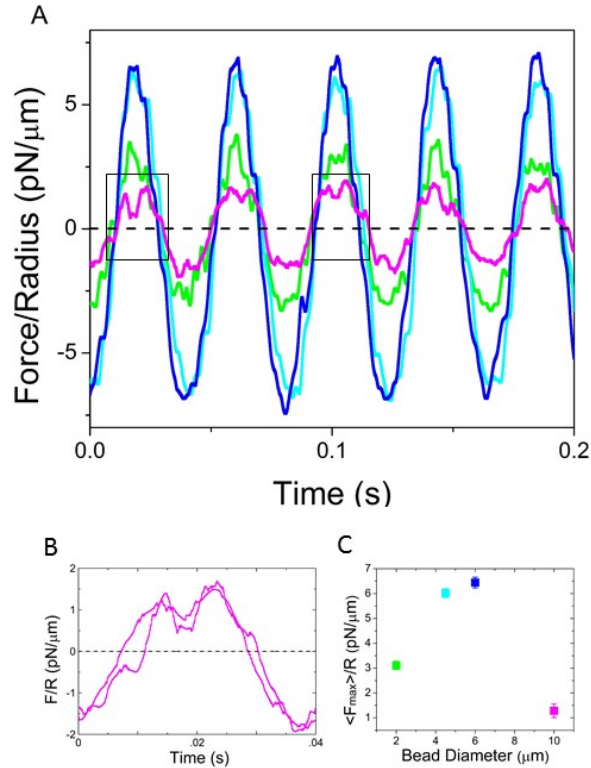

**Figure 2. Force response reveals lengthscale dependent material properties of mucus mesh**. (A) Sample force curves normalized by corresponding bead radii ($F/R$) show scale-dependent material response. Force curves shown are for an oscillation amplitude of 3.6 μm and frequency of 10 Hz for probe sizes of 2 (green), 4.5 (cyan), 6 (blue), and 10 μm (magenta). $F/R$ for 4.5 and 6 μm beads are nearly identical indicating the mucus is responding as a continuum fluid at these lengthscales ($l_2$). The ~2x reduction in $F/R$ for 2 μm probes indicates a transition to a water-like response where the bead is largely evading the mucus mesh ($l_1$). The ~3x reduced $F/R$ for the 10 μm probe, combined with the flat or dipped peaks (B), indicates that the bead is being forced out the trap, not tracking with the trap position during the complete oscillation. This forcing, unique to 10 μm probes, suggests the presence of more rigid structures at this lengthscale ($l_3$). (B) The force peaks from the boxed regions in (A) show that $F/R$ for 10 μm probes drop nearly to zero at the maximum position, as the bead is forced out of the trap. (C) The average normalized force amplitude, $<F_{max}>/R$, as a function of bead diameter clearly shows the described distinction between the three different lengthscales ($l_1$ (green), $l_2$ (cyan, blue), $l_3$ (magenta)).

The complex viscosity as a function of amplitude and probe size support this hypothesis (Figure 3). For 4.5 and 6 μm probes the viscosities are nearly identical, ~5x higher than that of water, for the entire amplitude range, suggesting that the mucus is responding as a continuum fluid at these size scales. The increased viscosity above that of water shows that the probes are indeed sensing the polymer network of the mucus. For 2 μm beads the viscosity is relatively constant for small amplitudes and is only ~2x the viscosity of water, indicating that the probe is traveling mainly through water, only rarely interacting with the mucus mesh. At an oscillation amplitude of ~4 μm there is a marked increase in the viscosity, approaching the corresponding values of the 4.5 and 6 μm beads, indicating a transition from a water-like regime to the continuum regime. At this and higher amplitudes, the 2 μm beads are moving far enough through the solution that they feel appreciable resistance from the polymer mesh that the larger probes move through. The increased variance in data at the higher amplitudes is also consistent with this picture as the heterogeneous nature of the mesh would lead to a wider spread in data than for a bead traveling through water.

If the mucus had a single correlation length that controlled the onset to continuum mechanics, then the viscosity for the 10 μm beads would be the same as that for the 4.5 and 6 μm beads. In contrast, we find that the viscosity measured with the 10 μm beads, is even lower than the 2 μm beads. This artificially low viscosity is an artifact of the resistive force of the mucus network forcing the bead out of the trap (resulting in low measured forces). Similar to the 2 μm beads, the apparent viscosity is constant up to ~4 μm after which it drops considerably because the probe can only travel a partial distance despite trying to move it further. Because $\eta \sim F_{max}/x_{max}$, if the measured force amplitude $F_{max}$ remains constant due to the bead being forced out of the trap at a specific strain distance by a rigid structure in the mucus, then as the amplitude $x_{max}$ increases, $\eta$ will decrease accordingly.

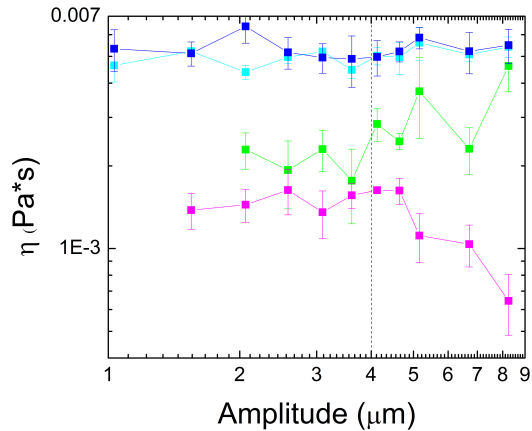

**Figure 3 Viscosity vs oscillation amplitude reveals a critical lengthscale of ~4 μm controlling mechanics**. Viscosity, $\eta$ (Pa-s), as a function of oscillation amplitude for 2 (green), 4.5 (cyan), 6 (blue) and 10 μm (magenta) probes. Note that the viscosity for 4.5 and 6 μm beads is nearly identical indicating a continuum fluid response at these scales ($l_2$). Smaller $\eta$ values for 2 μm probes indicate a transition to a water-like regime ($l_1$) which only approaches the continuum regime for oscillation amplitudes >4 μm ($l_2$, dashed line). The reduced $\eta$ for 10 μm probes is an artifact of the probe being forced out of the trap, likely by a more rigid structure in the mucus ($l_3$). The dashed line at an amplitude of ~4 μm indicates the point at which the mucus mesh begins to play a dominant role in mechanics ($l_2$).

Within this framework of a continuum fluid at the mesoscale ($l_2$) and a rigid scaffold at larger scales ($l_3$), we would expect the mucus to respond more elastically at larger scales, which we can determine by comparing the elastic modulus, $G'$, to the viscous modulus, $G''$ for each bead size (Figure 4). As expected, the elastic and viscous modulus for the 4.5 and 6 μm beads are nearly identical. Further, $G'' \approx 3G'$ over the entire amplitude range showing that the mucus response is largely viscous.

However, for 2 μm beads $G''$ is nearly and order of magnitude larger than $G'$, indicating that the 2 μm bead is not permeating the full network mesh and is instead sweeping largely through the water in which the mucus biopolymers are dissolved. However, there is a marked shift at ~4 μm in which $G'$ begins to increase with amplitude approaching the values measured for 4.5 and 6 μm beads. This increase is the source of the increase in the complex viscosity (Figure 3) and suggests that the fluid-like mesoscale mesh ($l_2$) is coupled to the macroscale elastic-like scaffold ($l_3$). In other words, because the probe is still smaller than the continuum lengthscale ($l_2$), the loosely entangled polymers could easily sweep past the probe as it moved through this mesoscale mesh unless it was coupled to the rigid scaffold. This coupling would provide elastic resistance that prevented the polymers from completely moving out of the way of the moving bead and provided an elastic restoring force. This can also be seen in the increased $G'$ for 4.5 and 6 μm beads relative to the water-like regime ($l_1$). This coupling is likely achieved via steric entanglements or chemical crosslinking as suggested to occur in other mucus systems (41, 43).

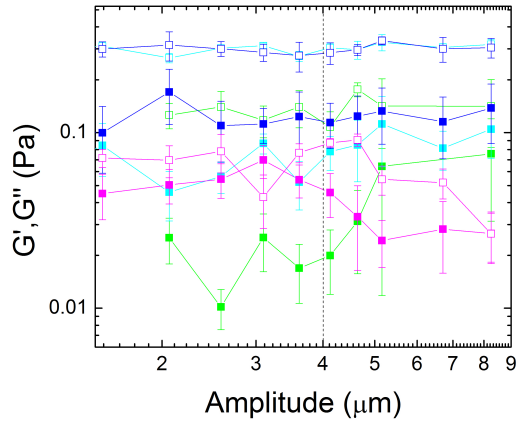

**Figure 4. Mesoscale polymer mesh at $l_2$ is highly viscous while elasticity is a largescale phenomenon controlled by larger rigid polymer scaffold at $l_3$.** Elastic modulus, $G'$, (closed squares) and viscous modulus, $G''$, (open squares) as a function of amplitude for four different probes sizes (colors are as in Fig 3). Note the similar values for the 4.5 and 6 μm beads and the increase in $G'$ for the 2 μm probe at $x_{max} > 4$ μm. Both features demonstrate a transition to a mesoscale continuum regime ($l_2$). The reduced gap between $G'$ and $G''$ for 10 μm beads demonstrates enhanced elasticity. The increase in $G'$ for 2 μm beads at $x_{max} > 4$ μm further supports a model in which the viscous continuum regime ($l_2$) is coupled to the largescale elastic network ($l_3$) forcing the polymers to collide with the 2 μm probe.

Finally, for 10 μm beads, $G'$ and $G''$ are comparable to each other indicating that elasticity is playing more of a role in the response at these larger size scales. Thus, the mucus mesh is not one simple entangled mesh with a single characteristic mesh size or entanglement spacing, but there may be two distinct networks, coupled to one another, that control the mucus mechanics in a lengthscale dependent manner. A similar "two-fluid" model has also been observed in human cervicovaginal mucus using electron microscopy (43). However, in these studies the material properties at these varying lengthscales were not characterized and coupling was not explored.

To further test this coupled multiscale mesh hypothesis and to characterize the perceived rigid elastic mesh we compare our results with macrorheology measurements. While our microrheology measurements demonstrate evidence of a rigid elastic network, the trap is not strong enough to hold the 10 μm bead and measure the full force response. By converting the microrheology amplitude values to strain ($x_{max}/2R$) we can quantitatively compare these measurements to macrorheology data (Figure 5) (56).

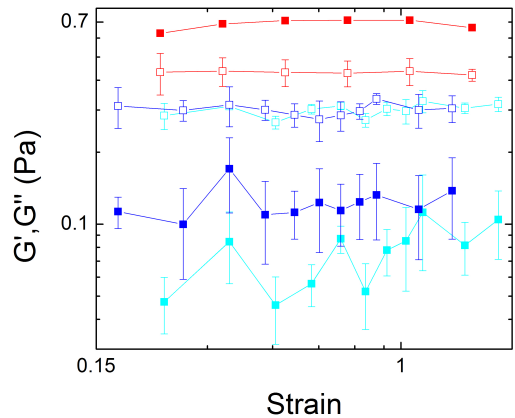

**Figure 5. Elasticity dominates the macroscale response, while viscosity is comparable on multiple lengthscales**. Elastic (closed) and viscous (open) modulus measured via macrorheology (red) and microrheology using 4.5 (cyan) and 6 μm (blue) microspheres. Macrorheology measurements exhibit $G' > G''$ (in contrast to microrheology data) indicating that elasticity is a macroscale phenomenon. $G''$ for both macrorheology and microrheology data are similar, indicating that the dissipative mechanics arise from the loosely entangled polymers in the mucus emerging at $l_2$.

For macrorheology measurements, $G'$ is larger than $G''$ for all measured strains, indicating that elasticity is a property revealed on the macroscale. Elastic gel-like networks are generally comprised of more rigid structures including bundled filaments that are highly entangled or crosslinked; and previous studies on mucins and human cervicovaginal mucus suggest that mucin polymers can self-assemble and cross-link into rigid bundled fibers (41, 43). Because each bundle is comprised of many mucins, the resulting mesh size for such a network of rigid bundles would be substantially larger than the mesh size of an entangled mucin network of the same concentration (41, 43). Thus, this macroscale elasticity likely arises from this larger, rigid mesh of bundled polymers that is probed by the 10 μm beads. Conversely, the macroscopic $G''$ is nearly the same as the microscale value for the 4.5 and 6 μm probes indicating that fluidity is the same at all scales and comes from the loosely entangled polymers that permeate the mucus and dominate the mesoscale mechanics ($l_2$).

To further characterize the lengthscale dependent material properties of the mucus we measure the frequency dependence of the viscoelastic moduli (Figure 6). For a network that exhibits little elasticity and flows largely like a Newtonian fluid (the terminal regime for entangled polymers), $G' \sim \omega^2$ and $G'' \sim \omega$. As elastic effects become more important, i.e. as the frequency of perturbations approach and surpass the relaxation or disentanglement rate of network, $G'$ approaches and surpasses $G''$ and both $G'$ and $G''$ initially approach frequency-independent plateaus. As shown in Figure 6, for 2 μm beads $G' \sim \omega^2$ and $G'' \sim \omega$ indicating that these probes interact very little with the entangled biopolymers in the mucus and travel mainly through the surrounding Newtownian fluid. For 4.5 and 6 μm beads, the scaling of $G'$ and $G''$ is slightly reduced from terminal regime scaling and the magnitudes of $G'$ and $G''$ are once again bead size independent, further demonstrating a mesoscale continuum regime. The reduced gap between $G'$ and $G''$ and the reduced scaling of $G'$ and $G''$ with frequency with these bead sizes relative to the 2 μm beads indicates a slightly more elastic regime from interacting principally with the polymer mesh. For 10 μm beads, $G'$ and $G''$ both approach frequency-independent plateaus with a further reduced gap between $G'$ and $G''$, indicating that these beads are coming into contact with a more rigid elastic mesh.

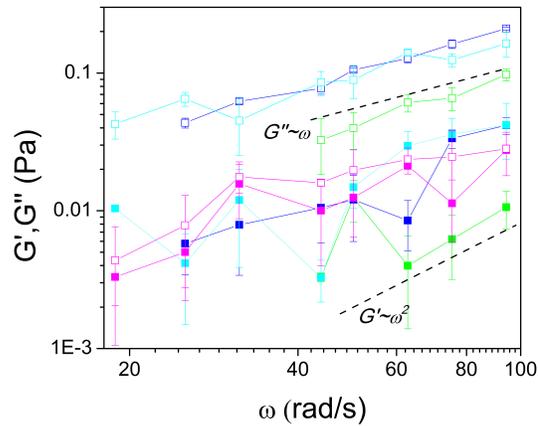

**Figure 6. Lengthscale dependent viscoelastic effects of mucus**. $G'(\omega)$ and $G''(\omega)$ for 2 (green), 4.5 (cyan), 6 (blue), and 10 μm (magenta) bead sizes. The scaling of $G'$ and $G''$ for 2 μm beads indicates that the mucus is responding principally as a Newtonian fluid with minimal elasticity, indicated by the terminal regime scaling laws represented by dashed lines. For the larger probes, the scaling of both $G'$ and $G''$ deviate from terminal regime scaling (dashed lines) indicating the onset of viscoelastic effects from interacting with the entangled mucus mesh. Consistent with our proposed model, the values for 4.5 and 6 μm beads coincide ($l_2$), and for 10 μm beads $G'$ approaches $G''$ (enhanced elasticity) and both $G'$ and $G''$ approach frequency-independent plateaus (enhanced elasticity).

We quantify the relative elasticity of the mucus as a function of probe size by comparing the average ratio of the elastic modulus to viscous modulus $<G'/G''>$ for all four bead sizes as well as for macrorheology. This quantity can also be understood as the inverse of the loss tangent, $\tan\delta = \sin\Delta\phi/\cos\Delta\phi$, which quantifies the relative dissipation in the system. Figure 7 shows that the relative elasticity increases with increasing bead size, and for 10 µm beads this ratio is comparable to the ratio of the macroscopic moduli, demonstrating that the rigid mesh responsible for pulling the 10 µm bead out of the trap is also what leads to the macroscale elastic response. Examining $G'/G''$ as a function of amplitude demonstrates that the elastic component of the mucus begins to play a role at a lengthscale of ~4 µm, indicated by $G'/G''$ values for all bead sizes converging. This convergence demonstrates that the mesoscale mesh at $l_2$ is indeed likely coupled to the largescale ($l_3$) scaffold responsible for the elastic contribution to mechanics.

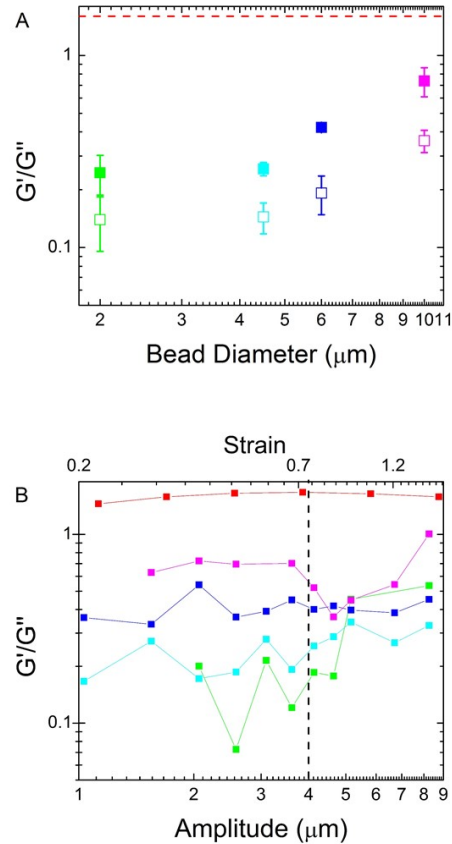

**Figure 7. Relative elasticity $<G'/G''>$ of mucus increases with increasing lengthscales.** (A) Average relative elasticity $<G'/G''>$ for 2 (green), 4.5 (cyan), 6 (blue), 10 µm (magenta) beads and macrorheology data (red). Closed squares are $<G'/G''>$ values averaged over all amplitudes while open squares are averaged over all frequencies. The relative elasticity increases as bead size increases, approaching $<G'/G''>$ for macrorheology data, indicating that elasticity is a macroscale phenomenon. (B) $<G'/G''>$ as a function of amplitude shows that elastic effects begin to play a role in mechanics beyond 4 µm, indicating that the mesoscale polymer mesh ($l_2$) is coupled to the rigid scaffold ($l_3$).

**Conclusion:**

We have used active microrheology to determine three intrinsic lengthscales governing the structural properties of *Chaetopterus* mucus, and characterize the corresponding length-scale dependent material properties. Below ~4 μm ($l_1$) the mucus responds similarly to water with minimal contribution from the mucus mesh. For lengthscales between 4-10 μm ($l_2$) a mesoscale continuum regime emerges with principally viscous response features characteristic of a loosely entangled polymer mesh. For lengthscales ≥10 μm ($l_3$) the mucus responds elastically and mimics the macroscopic mucus properties. These collective results suggest that the mucus can be modeled as a coupled two-fluid system comprised of (i) a loosely entangled polymer mesh with pore sizes of ~4 μm coupled to (ii) a larger more rigid scaffold with pore sizes of ~10 μm (Figure 8). While mesh (i) is responsible for the enhanced viscosity of the mucus at lengthscales above $l_1$, mesh (ii) provides the macroscale elasticity of the mucus at $l_3$.

This model can explain how mucus is able to simultaneously perform a multitude of functions across varying lengthscales. For example, $l_1$ is necessary for the passage of nutrients while $l_2$ is needed for trapping larger pathogens. Finally, the macroscopic gel-like properties necessary for coating organs is provided by $l_3$. Findings can also aid in developing emerging targeted drug delivery systems in which nano- or microscopic particles must either pass through or become trapped by the mucus.

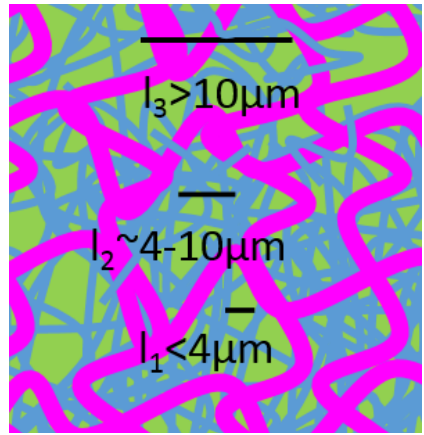

**Figure 8. Microrheology measurements suggest that the mucus is comprised of a complex biopolymer mesh with three intrinsic lengthscales**. Cartoon of proposed structure of mucus. $l_1$ is the water-like regime, where the mesh plays little role in the response and particles can freely pass through the mucus. $l_2$ is the intermediate mesoscale continuum regime comprised of loosely entangled polymers; here the mucus can be treated as a single homogeneous material. $l_3$ is the regime where the rigid scaffold, which produces the elastic-like macroscopic mucus properties, dominates mechanics. Coupling of the two meshes at $l_2$ and $l_3$ is likely achieved via steric entanglements or chemical crosslinking.

**Acknowledgements:**

We are grateful to Marine Collector Phil Zerofski at Scripps Institution of Oceanography for collecting *Chaetopterus* worm specimen, Dr. Cole Chapman (USD) for software development and preliminary data collection, and undergraduate student Jenny Tu (SIO) for chemical analyses. This work was supported by the Air Force Office of Scientific Research Young Investigator Program, Grant No. FA95550-12-1-0315 (RMRA), as well as Grant No. FA9550-14-1-0008 (DDD) and FA9550-14-1-0171 (JSU and DLB). It was also the subject of a travel grant from the Burroughs Wellcome Fund (Grant No. 1012729) for DDD to perform analyses at Georgetown University.